\documentclass[12pt]{iopart}
\usepackage{graphics,graphicx}
\usepackage{dcolumn,bm}
\usepackage{psfrag}
\usepackage{color}
\usepackage{multirow}

\begin{document}

\title{$A+ A \to \emptyset$ reaction for particles with a dynamic bias   to
move away from their nearest neighbour in one dimension}
%[$A+ A \to \emptyset$ reaction with dynamic bias]

\author{$^1$Reshmi Roy and $^2$Parongama Sen}
\address{$^{1,2}$ Department of Physics, University of Calcutta,
92 Acharya Prafulla Chandra Road, Kolkata 700009, India.}

\ead{$^1$reshmi.roy80@gmail.com}
\vspace{10pt}

\begin{abstract}
We consider the dynamics of particles undergoing the reaction   $A+A \to \emptyset$  in one dimension with a dynamic bias.
Here the particles move towards
their nearest neighbour with probability $0.5+\epsilon$ where $-0.5 \leq \epsilon < 0$.
$\epsilon_c = -0.5$ is the  deterministic limit where the nearest neighbour interaction is strictly repulsive. 
We show that the negative bias changes drastically the behaviour of the  
fraction of surviving particles $\rho(t)$ and  
persistence probability $P(t)$ with time $t$. 
$\rho(t)$ decays as $a/ (\log t)^b$ where   $b$ increases with  $\epsilon - \epsilon_c$.
$P(t)$  shows a stretched exponential decay with non-universal decay  parameters.
The probability 
$\Pi(x,t)$ 
that a tagged particle is at position $x$ from its origin  
is found to be Gaussian for all $\epsilon<0$; 
 the associated scaling variable is $x/t^\alpha$ where  $\alpha$  approaches  the 
known limiting value $1/4$ as $\epsilon \to  \epsilon_c$, in a power law manner. 
Some additional features of the dynamics by tagging the particles are also studied.  The results are compared to the case of positive bias, a well studied problem. 
\end{abstract}

%\maketitle
\section{Introduction}
Reaction diffusion systems have been extensively studied over the last few decades, especially in one dimension  %
\cite{privman,ligget,krapivsky,odor,derrida_95,racz,amar,avraham,alcaraz,krebs,santos,schutz,oliveira}. 
The simplest form of a reaction diffusion system is 
$A+A\ \to \emptyset$, where the particles $A$ diffuse and annihilate on contact. 
This model in one dimension, with asynchronous updating, also represents the ordering dynamics
of the Ising model with Glauber dynamics at zero temperature. 
%This system can
%describe various physical, biological or social problems e.g., pattern formation and opinion formation in
%society etc. 
%In the lattice representation of this model, 
When considered on a lattice, one can assume that the  particles $A$ occupy the sites of the  lattice
and at each time step they hop to a nearest neighbouring site.

The $A+A \to \emptyset$ system has been studied in the recent past where the particles $A$ move with a bias  
towards their nearest neighbours \cite{soham_ray_ps2011, ray_ps2015,roy2020} in one dimension.  The model, in its deterministic limit,  maps to a opinion dynamics model studied earlier \cite{soham_ps2009}. 
Previously,   both the bulk dynamical and  
tagged particle dynamics have been  reported in the one dimensional $A+A \to \emptyset$ system where the particle $A$ diffuses towards its 
nearest neighbour with a probability $0.5+\epsilon$ $(0<\epsilon\leq 0.5)$ and in the opposite 
direction with probability $0.5-\epsilon$. The results show significant differences when compared to the case with no bias ($\epsilon =0$) although  the annihilation process is identical in the latter.  
This reaction diffusion model with parallel updating has also been studied in two dimensions recently \cite{pratik_ps2019}. 

%The annihilation process is not affected 
%by the bias but this extension leads to drastic changes in bulk dynamical 
%properties. 

To generalize the problem, in the present paper, the results for a negative bias are reported, i.e.,  
when  $\epsilon < 0$.  
The idea behind the study is to find the universal behaviour in the bulk properties as well as the 
microscopic features. Here we have used asynchronous dynamics to compare with the positive bias case results
which have already been studied before.
 Specifically, $\epsilon=-0.5$ implies purely repulsive motion where the particles always 
move towards their farther neighbour. 
These particles with full negative bias can represent 
the motion of similarly charged particles or in general particles with repulsive interaction which  can move both ways. Henceforth we denote the fully biased point $\epsilon = -0.5$ by $\epsilon_c$.

\section{The  Model, dynamics and simulation details} \label{sec_model}

\begin{center}
\begin{figure}[h]
\includegraphics[width= 15cm]{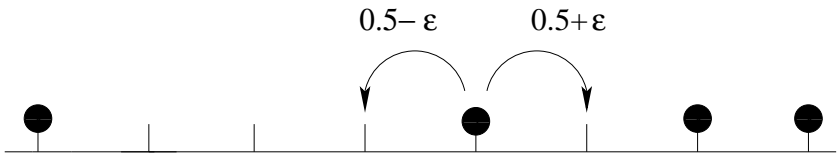}
\caption{The tagged particle hops to right with probability 
$0.5+\epsilon$ as its right neighbour is two lattice separation away and 
to left with probability $0.5-\epsilon$ as the left neighbour is four lattice
separation away. In the present case as $\epsilon<0$, the particle has a preference to
move in the left direction as its nearest neighbour is on its right. For  
$\epsilon=-0.5$, the particle definitely moves to the left. In comparison, in the conventional reaction diffusion case, $\epsilon =0$ and the particle has equal probability to move left and right.}
\label{model}
\end{figure}
\end{center}

In the $ A+A \to \emptyset$ model, a particle $A$ diffuses to one of its neighbouring
sites and undergoes a reaction (annihilation).  Here, at each update, a site is randomly chosen and if there is a 
particle
on the selected site, it hops one step towards its nearest neighbour with probability
 $0.5+\epsilon$ (and with probability $0.5-\epsilon$  in the opposite direction);  $-0.5 \leq \epsilon<0$. 
If the destination site is previously occupied by a particle, both of them will be annihilated 
simultaneously. The position  of the particle is updated immediately in the asynchronous scheme of updating. 
 $L$ such updates constitute one Monte Carlo step (MCS).
 In the rare cases of two equidistant neighbours, the particle
moves in either direction with equal probability 0.5. The motion
is illustrated in figure \ref{model}. 
It may be noted that the direction of motion is determined by the relative distances of the neighbouring particles only;
the particle has a tendency to move away from the nearest neighbour (for $\epsilon < 0$, which is the choice here). 
The actual distances are of no consideration in the present scenario. 
Also, if a site is chosen for updating, the particle sitting at that
site has to perform a move. 
%However, as sites are chosen randomly, it may happen that in one Monte Carlo time step, the position 
%of a particle is not updated at all. Also, since asymchronous dynamics are being used, the particle positions are updated immediately, and 
%this may happen more than once within a Monte Carlo step. The effect of this will be manifested in certain quantities, discussed later in the 
%ms. 

As asynchronous dynamics have been used, there are several interesting points to be noted. The net displacement of a particle
can be zero or more than one after the completion of one  MCS \cite{roy2020}. 
This affects the numerical estimates of certain quantities 
that have been estimated in the present work. 
For the fully biased point $\epsilon_c$, annihilation can occur only if 
three particles occupy immediately adjacent sites, however, in the asynchronous update scheme, whether an annihilation will take place will 
depend on which site is getting updated first, so it is a necessary but not sufficient condition.

The studies are performed on lattices of maximum size $L = 24000$ and the maximum
number of initial configurations taken  is  2000. Periodic boundary condition has
been used in all the simulations.
We have considered the lattice of size $L$ to be randomly half filled initially.

\section{Simulation  Results} \label{results}

\begin{center}
\begin{figure}[h]
\includegraphics[width= 15cm]{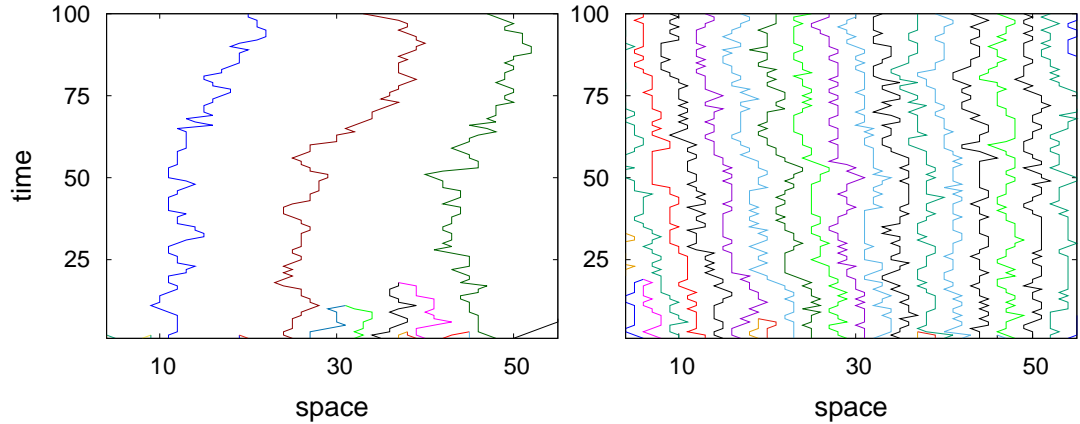}
\caption{ Snapshots of the system at different times for $\epsilon=-0.1$
(left) and $\epsilon=-0.5$ (right).
The snapshots are shown for a  part of a system of size $L=200$. The
trajectories of different particles are represented by
different colours.}
\label{snapshots}
\end{figure} 
\end{center}

We took snapshots of the system to check the motion of individual 
particles. The world lines of the motion of the particles are shown 
for $\epsilon=-0.1$ and $-0.5$ in figure \ref{snapshots}.
It  may be noted immediately  they are 
strikingly different from each other. It is obvious that the number of
annihilation is larger for $\epsilon=-0.1$ and it is left with much fewer
particles within the same timescale. Also, the paths traced out resemble more a diffusive 
trajectory.  In contrast, for $\epsilon = -0.5$, the particles change their direction more often 
and remain confined within a  limited region in  space. 

To probe the dynamics of the particles, we have studied the
following quantities: (i) fraction of surviving particles $\rho(t)$ at time $t$,
 (ii) persistence probability of the lattice sites $P(t)$,
(iii) the probability distribution 
$\Pi(x, t)$ of finding a particle $A$ at distance $x$ from its
origin at time $t$, (iv) the probability $S(t)$ of the change in
the direction in the motion of a particle at time $t$ and 
(v) the distribution $D(\tau)$ of the time interval
$\tau$ between two successive changes in the direction of the
motion of a particle. The results for each of these quantities 
are presented in the following subsections.

\subsection{Bulk properties}

\subsubsection {Fraction of surviving particles $\rho(t)$}
 
\begin{center}
\begin{figure}[h]
\includegraphics[width= 9cm]{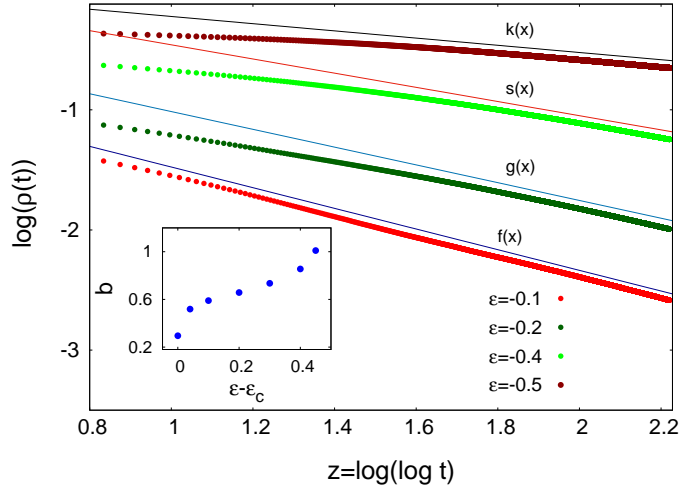}
\caption{ Variation of the fraction of surviving particles $\rho(t)$ with  time $t$
is studied by plotting   $\log(\rho(t))$ against $z = \log(\log t)$ and the best fit lines (according to equation (\ref{aliveeq})) along with, shifted vertically for better visualisation.   
The best fit lines are (a) $f(z)=\log(0.51)-0.86z$ for $\epsilon=-0.1$, (b) $g(z)=\log(0.7)-0.74z$ for $\epsilon=-0.2$,
(c) $s(z)=\log(1.07)-0.59z$ for $\epsilon=-0.4$ and
(d) $k(z)=\log(1.01)-0.3z$ for $\epsilon=-0.5$. 
 These results are for a system size $L = 8000$.
 Inset shows the variation of $b$ with $\epsilon-\epsilon_c$, where $\epsilon_c=-0.5$. The errors are less than the size of the data points.
%The maximum error in $b$ obtained from fitting $\rho(t)$  to the form given in equation (\ref{aliveeq}) is typically of the order of 0.03$\%$.

}
\label{alive_asyn}
\end{figure}
\end{center}

For the purely diffusive system ($\epsilon=0$), it is well known that the fraction of surviving
particles shows a power law behaviour in time; $\rho(t)\sim t^{-\gamma}$ with $\gamma=0.5$.
If a positive bias in introduced in the system, $\gamma \approx 1$ for all $\epsilon>0$ \cite{soham_ray_ps2011,ray_ps2015}. 
The exponent increases as the attractive dynamics result in an increased number of annihilation. 
As $\epsilon$ is made negative, the number of annihilation decreases as reaction becomes less probable
because of the repulsion.
So, $\rho(t)$ shows a slow decay in time
and can be fitted to the following form 
\begin{equation}  
\rho(t)=a/{(\log t)}^b,
\label{aliveeq}
\end{equation}  
where $a$ and $b$ are constants, depending on $\epsilon$. The fitting is made with a two parameter least square fitting using GNUFIT. In figure \ref{alive_asyn}, $\log \rho(t)$ is plotted against $\log(\log t)$ 
for different $\epsilon$ values,
to 
manifest the linear dependence at long times. 
%However, a weak  deviation from linearity is noted  at very large times  which %suggests there may be a correction term to the above variation}.    
Here it may be mentioned that for 
the extreme point  $\epsilon=\epsilon_c$,  the particles ideally attain a  equidistant configuration. But the  dynamical
rule is such that the particles have to make a move and hence they perform  a nearly oscillatory motion.  Annihilation
takes place extremely rarely at large times such that $b$ decreases as the 
magnitude of $\epsilon$ increases 
 (see inset of figure \ref{alive_asyn}). 
%Hence $b$ shows a decrease with $\epsilon$. 

\subsubsection{Persistence probability $P(t)$}

\begin{center}
\begin{figure}[h]
\includegraphics[width= 10cm]{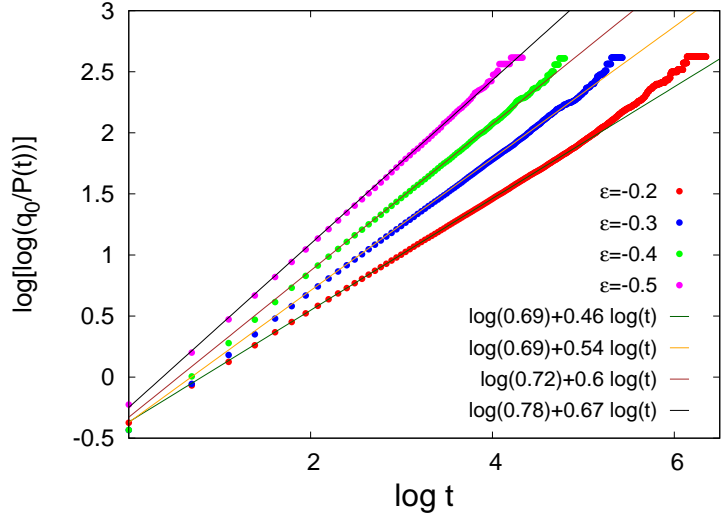}
\caption{Variation of the persistence probability $P(t)$ with time $t$ is studied by plotting $\log(\log\frac{q_0}{p(t)})$ against $\log t$  for several $\epsilon$ and 
the best fit lines (according to equation (\ref{pereq})) are shown along with for different $\epsilon$ in the same order. These results are for a system size $L=8000$.}
\label{persis}
\end{figure}
\end{center}

Persistence probability $P(t)$ in this model is defined as the probability that a site is unvisited till time $t$.
For $\epsilon=0$, $P(t)$ decays as $P(t) \sim t^{-\theta}$ with ${\theta} = 0.375$ \cite{derrida}. For $\epsilon>0$,
 $\theta \approx 0.235$, however small be the bias \cite{soham_ray_ps2011}.
As $\epsilon$ becomes negative, $P(t)$ falls off rapidly (see figure \ref{persis}). $P(t)$ shows a stretched exponential decay in time:
\begin{equation}
P(t)=q_0\exp(-qt^r).
\label{pereq}
\end{equation}
Once again, the best fit curves with a three parameter function 
are obtained using GNUFIT. In figure \ref{persis} we show the validity of the above form by obtaining linear dependence when $\log (\log q_0/ P(t)) $ is plotted against $\log t$. 
Both $q$ and $r$ increase as $\epsilon \to \epsilon_c$.

\subsection{Tagged particle features}

\subsubsection {Probability distribution $\Pi(x, t)$}

For pure random walk ($\epsilon=0$), the probability distribution $\Pi(x,t)$ 
is known to be Gaussian and $\Pi(x,t)t^{1/2}$ shows a data collapse for
different times when plotted against $x/t^{1/2}$. This is also true for the 
unbiased ($\epsilon = 0$) annihilating random walkers 
 because they perform purely diffusive motion until they are annihilated. For $\epsilon<0$,
 the distributions can again be fit to a Gaussian form. 
%(see figure \ref{prob_distri}). 
However the scaling variable is in general $x/t^\alpha$ with $\alpha<0.5$.
%To obtain data collapse for different times we have plotted
%$\Pi(x,t)t^\alpha$ against $x/t^\alpha$. 
We extract the value of $\alpha$ from the data using two different methods.

\begin{center}
\begin{figure}[h]
\includegraphics[width=16cm]{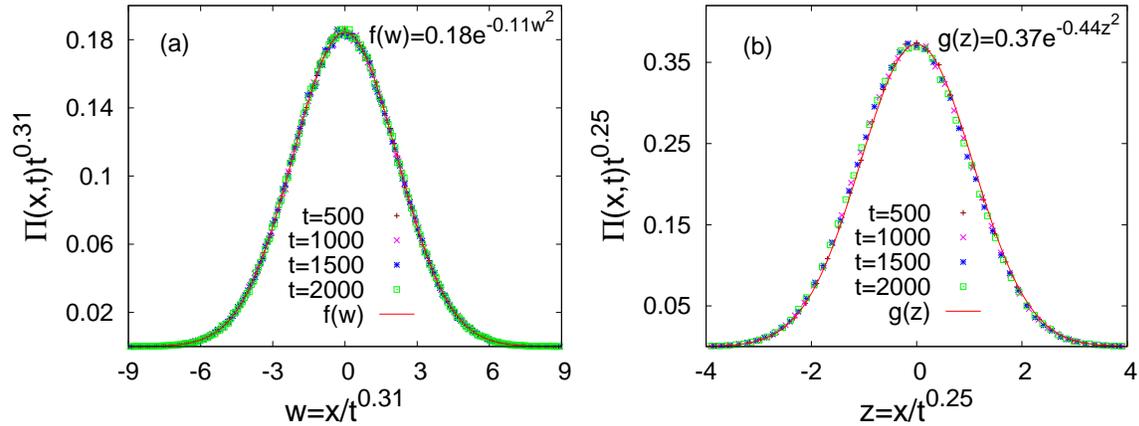}
\caption{Data collapse of probability distribution $\Pi(x,t)$ is studied by plotting $\Pi(x,t)t^\alpha$ against $x/t^\alpha$ for $\epsilon=-0.1$ (a) and $\epsilon=-0.5$ (b), where $\alpha=0.31$ for $\epsilon=-0.1$ and $\alpha=0.25$ for $\epsilon=-0.5$, obtained using Method I for a system of size $L=12000$. The collapsed data are fitted to the Gaussian distribution functions $f(w)$ and $g(z)$ for $\epsilon=-0.1$ and $\epsilon=-0.5$ respectively.}
\label{prob_distri}
\end{figure} 
\end{center}

\begin{center}
\begin{figure}[h]
\includegraphics[width= 12cm]{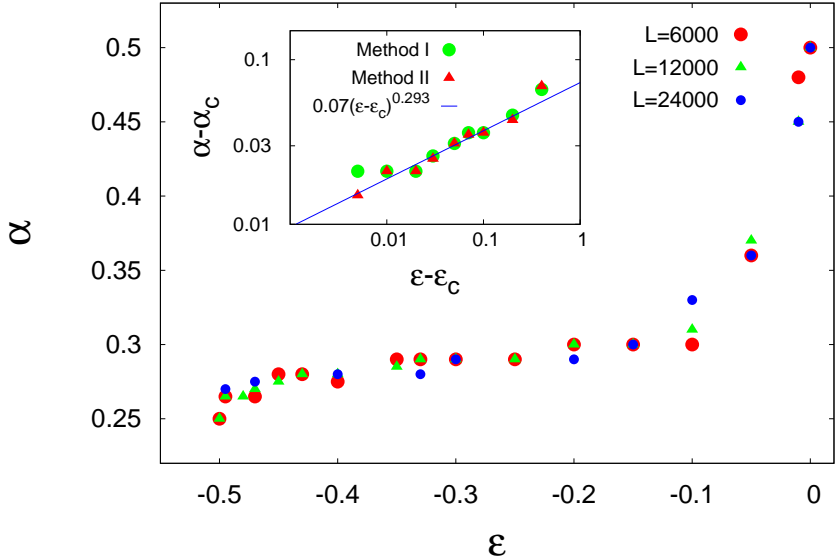}
\caption{Variation of $\alpha$ is shown with $\epsilon$ for several system sizes using Method I. Inset shows $\alpha - \alpha_c$ against $\epsilon - \epsilon_c$ for Methods I and II for a system size $L=12000$ where $\alpha_c=0.244 \pm 0.001$ corresponds to $\epsilon=\epsilon_c= -0.5$ and a power law fitting is shown for the values obtained from Method II.
The errors involved in the estimate of $\alpha$ from Method II is $\pm 0.001$ while for Method I it is typically $0.005$.}

\label{scale_asyn}
\end{figure}
\end{center}

 Method I: In this method the scaling variable $\alpha$ is obtained by collapsing the 
data using  trial values of $\alpha$  and choosing the value for which the data collapse looks most impressive
(see figure \ref{prob_distri}). 
This  analysis   indicates  that     
$\alpha$ depends on  $\epsilon$, the values are 
 shown in figure \ref{scale_asyn}. There seems to be some  finite size
dependence which, however,  could not be systematically captured  in this method. 
%$\alpha$ deviates from 0.5 and decreases as $\epsilon$ becomes negative.
As $\epsilon$ decreases from zero, at first $\alpha$ decays rapidly from the value 0.5 
at $\epsilon=0$ until $\epsilon \simeq - 0.1$ where it attains a value close to 0.3. 
Below $\epsilon = -0.1$, $\alpha$ shows a slow decrease and at $\epsilon_c$,
it is close to  0.25, the value expected for repulsive random walkers \cite{arratia}. 
At large time as the walkers do not annihilate, effectively they perform repulsive
random walk in the lattice.   
The typical  error involved in the above estimates is  $\pm  0.005$.

Method II: The values of $\alpha$ obtained from Method I indicates that $\alpha$ has a comparatively weaker variation with $\epsilon$ for $\epsilon < -0.1$. 
 To  obtain a more accurate dependence of $\alpha$   for $\epsilon < -0.1$, we employ  
another  method  that optimises the value of $\alpha$ needed to obtain the best data collapse.  Here we utilise the fact that the scaling function is  Gaussian.  Method II is based on  the prescription given in  \cite{somen}, when the form of the scaling function is known. 

For a given $\epsilon$ value, we use the  same four sets of data 
corresponding to four different times that were 
used to get the collapse  in Method I.  
We first choose a value of $\alpha$ and taking any of the four sets of data (say, the $i$th set with a probability distribution  $\Pi_i (x,t)$),   fit a Gaussian function to the scaled probability distribution $\Pi_i (x,t)t^\alpha$. The    scaling  variable here is $x/t^\alpha$
 such that  
\begin{equation}
\Pi_i(x,t)t^\alpha = a_i \exp[-b_i(x/t^\alpha)^2].
\end{equation}
Knowing   $a_i$ and $b_i$  from the  fitting, we
now choose another set $j \neq i$, and  estimate the deviation from the above  Gaussian 
function by calculating 
\begin{equation}
e_{j}^{i} = \langle \bigg[\bigg(\Pi_j(x,t)t^\alpha - a_i \exp[-b_i(x/t^\alpha)^2]\bigg)^2\bigg] \rangle,
\end{equation}
$\langle....\rangle$ denotes average over all the discrete points $x/t^\alpha$ in the $j$th data set.
The total averaged error for the choice of the $i$th set as the initial set is then equal to $ E_i = \sum_{j\neq i} e_j^i/3$.

Next we repeat the above exercise by choosing a different set as the initial 
set to get $E_i, ~ i=1,2,3, {\rm {and}}~ 4$ and finally compute the averaged error 

\begin{equation}
E(\alpha) = \frac{1}{4} \sum_{i=1}^{4} E_i.
\end{equation}
Plotting $E(\alpha)$ against  $\alpha$, a minimum value is expected at a certain value of $\alpha$ which is identified as the optimal value
that gives the best data collapse.

\begin{center}
\begin{figure}[h]
\includegraphics[width= 16cm]{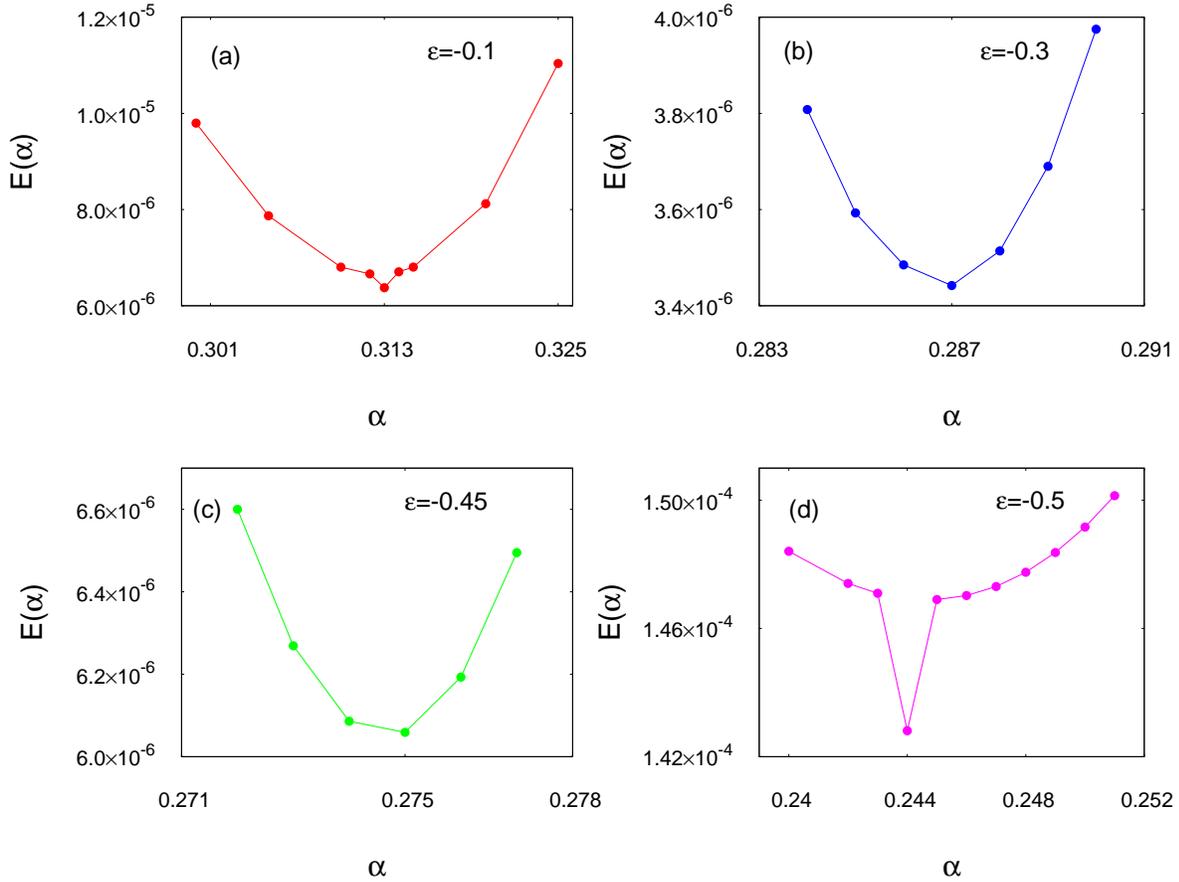}
\caption{Variation of  $E(\alpha)$ for different values of  $\epsilon$ for $L=12000$.}
\label{e-alpha}
\end{figure}
\end{center}
 
A minimum value of $E(\alpha$) is indeed obtained   as 
we vary the value of $\alpha$ in steps of 0.001. 
  The results for $E(\alpha)$ are shown for 
different vales of $\epsilon$ in figure \ref{e-alpha}. The values of $\alpha -\alpha_c$ where $\alpha_c $ corresponds to $\epsilon = \epsilon_c$  are plotted in the inset of figure \ref{scale_asyn} obtained from both the methods and a log-log plot shows that a variation 
\begin{equation}
\alpha - \alpha_c \propto |\epsilon_c -\epsilon|^{0.293\pm 0.029} 
\end{equation}
is quite compatible with the more accurate estimates of Method II close to 
small values of $\epsilon-\epsilon_c$ .

\subsubsection {Probability of direction change $S(t)$}

\begin{center}
\begin{figure}[h]
\includegraphics[width= 12cm]{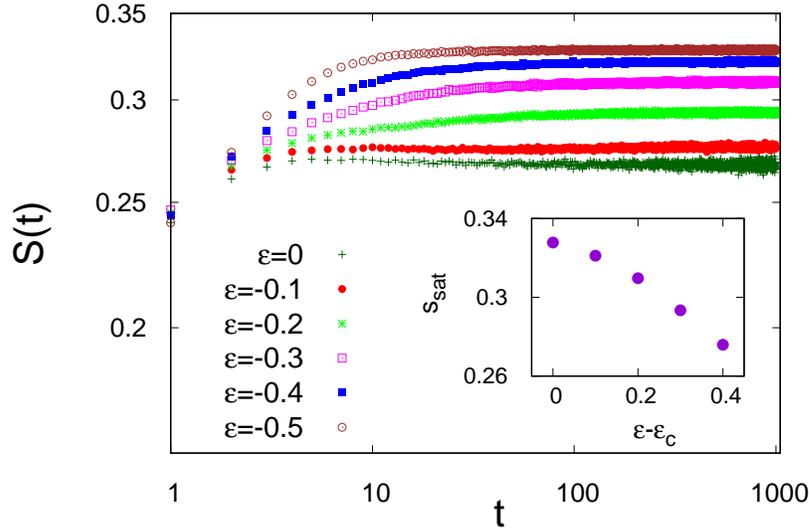}
\caption{Probability of direction change $S(t)$ of a tagged particle at time $t$ for
different $\epsilon$ for $L=10000$. Inset shows variation of $S_{sat}$
with $\epsilon-\epsilon_c$. The errors are less than the size of the data points}.
\label{dirchange}
\end{figure}
\end{center}

The probability of direction change of a particle is calculated  by estimating the number of particles 
that changes direction of motion at time $t$ divided by the number of surviving particles at that time.
 Figure \ref{dirchange} shows the data for $S(t)$ for different
$\epsilon$. For purely diffusive system $(\epsilon=0)$, $S(t)$ is 
independent of time, $S(t)=p_0$. $p_0$ is dependent on the dynamical 
updating rule, it turns out to be $\sim 0.27$  numerically 
with the asynchronous updating  rule used here \cite{roy2020}.  

For $\epsilon<0$, at first $S(t)$ increases with time, then it reaches a constant value $S_{sat}$.
Repulsion between the neighbouring particles is mainly responsible for the change in direction of motion.
When $\epsilon$ decreases from zero the repulsive factor becomes stronger, particles
change their direction more rapidly, $S(t)$ increases. 
At the extreme limit $\epsilon=\epsilon_c$, the change in direction is maximum as the particles 
perform nearly oscillatory motion. A systematic decrease of the saturation value 
is obtained when $S_{sat}$ (calculated from the last 500 steps) are plotted against $\epsilon-\epsilon_c$
(see inset of figure \ref{dirchange}).

\subsubsection{Distribution of time interval spent without change in direction of motion $D(\tau)$}
Another quantity calculated is $D(\tau)$,  the probability distribution of  the interval of
time $\tau$ spent in between two successive changes in direction. 
 A particle may continue to move in the same direction for different intervals of time denoted by $\tau$. For each tagged 
particle, these intervals   are calculated (up to a particular time $t$) to obtain the  distribution $D(\tau)$ at $t$, 
which is normalised such that $\sum_\tau D(\tau) = 1$.  Here we have calculated $D(\tau)$ at  $t=1000$. 

\begin{center}
\begin{figure}[h]
\includegraphics[width= 10cm]{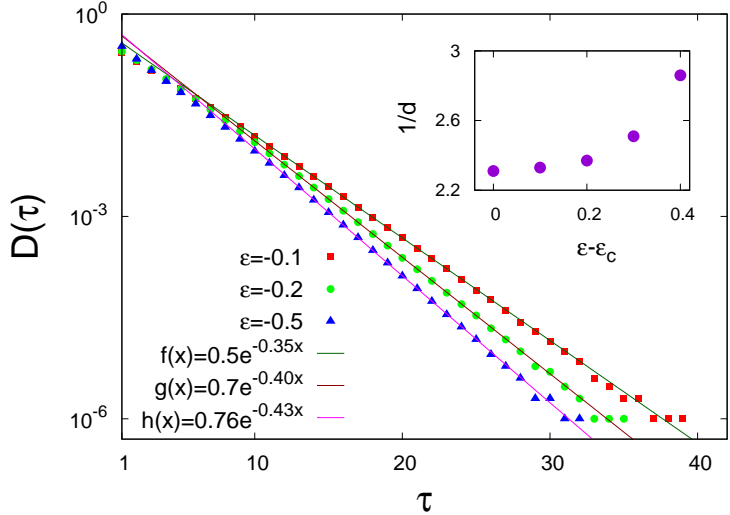}
\caption{Variation of $D(\tau)$ over $\tau$ is studied at time $t=1000$ and shown in a log linear plot for several $\epsilon$. The best fit lines (according to equation
(\ref{timedisteq})) are shown along with for different $\epsilon$ in the same order. Inset shows the variation of $\frac{1}{d}$ with $\epsilon-\epsilon_c$.
 The errors are less than the size of the data points. These results are for system size $L=10000$.}
\label{timedist}
\end{figure}
\end{center}

For
random walkers with $\epsilon=0$, 
$D(\tau$) is given by
\begin{eqnarray}
D(\tau)={p_0}^{2}({1-p_0})^{\tau},
\label{timeeq}
\end{eqnarray}
which reduces to an exponential form: $D(\tau)\propto \exp[-\tau \ln \{1/(1-p_0)\}]$. 
As for $\epsilon<0$,   $S(t)$ is a constant at large times, $D(\tau)$ is expected to
show an exponential decay. Therefore,  $D(\tau)$  is fitted 
according to 
\begin{eqnarray}
D(\tau)=c\exp(-d\tau).
\label{timedisteq}
\end{eqnarray} 
Figure \ref{timedist} shows the data for $D(\tau)$ against $\tau$ for different values of  $\epsilon$ calculated at time $t=1000$. 
$1/d$ is an effective `time scale' which increases with $\epsilon-\epsilon_c$, shown in the
inset of figure \ref{timedist} (calculated from the tail of the distribution). It shows that for $\epsilon_c$, the tendency to
oscillate is maximum.
%As $\epsilon$ decreases $S(t)$ increases, $q$ also increases (see inset of figure \ref{timedist} ). 

In principle, the value of $d$ in equation (\ref{timedisteq}) should be identical to 
$\ln\{1/(1-S_{sat})\}$. In order to check this, a careful 
inspection of the behaviour of $D(\tau)$ shows that $d$ has a different value 
for  small $\tau$ (up to $\tau \approx   15$)  and for  the tail of the distribution.
We have tabulated both the values obtained from the two regimes in Table \ref{table1}  
as well as 
the values of $\ln\{1/(1-S_{sat})\}$ for comparison.  Evidently, the values of the latter quantity match better with  the $d$ values obtained from the smaller $\tau$ region of $D(\tau)$. This may be because for  larger $\tau$, the statistics is poorer as $D(\tau)$ follows an  exponential distribution. The discrepancy between the calculated value of $d$ from $S_{sat}$ and that from large $\tau$ region of $D(\tau)$  increases systematically with the magnitude of $\epsilon$ which may be because we are calculating $D(\tau$) at the same time 
for all $\epsilon$.

%  Due to the presence of fluctuation in $S(t)$, an error is also involved 
% in the calculation. For instance, $S_{sat}=0.320$ and $\Delta S_{sat}=0.005$ for $\epsilon=-0.4$, this results in $d=0.385 \pm 0.007$. On the other hand, the error in $d$ obtained from fitting $D(\tau)$  to the form given in equation (\ref{timedisteq}) is typically of the order of 0.1$\%$.

\begin {table}[h]
\caption{Comparison of $d$ with $\ln \frac{1}{1-S_{sat}}$; typical errors  are of the order of 1$\%$ for all the  estimates.}
\begin{center}
\begin{tabular}{ |c|c|c|c|c| }
\hline
$\epsilon$ & $S_{sat}$& $\ln \frac{1}{1-S_{sat}}$ & $d$ (for small $\tau$) & $d$ (for large $\tau$) \\
\hline
\hspace{1mm}  -0.1 \hspace{1mm}   &  \hspace{1mm}   0.276 \hspace{1mm}  &  \hspace{1mm}   0.323 \hspace{1mm}   &   0.320   &  0.350  \\
\hline
-0.2    &  0.293   &    0.347   &   0.346 &  0.400   \\
\hline
-0.3    &  0.310   &    0.371   &  0.370  &  0.423    \\
\hline
-0.4    &  0.321   &     0.387   &  0.383 &  0.430      \\
\hline
-0.5    &  0.328   &   0.396   &   0.397  &  0.431     \\
\hline
\end{tabular}
\end{center}
\label{table1}
\end {table}

\section {Concluding remarks}

In this paper, we have studied the behaviour of the $A+A \to \emptyset$ model in one dimension, where
the particles tend to avoid their nearest neighbour. The probability to move towards the nearest neighbour is taken
parametrically as $0.5 + \epsilon$ where $\epsilon < 0$. The case with $\epsilon > 0$ has been 
studied earlier \cite{soham_ray_ps2011,ray_ps2015,roy2020}. 
The  bulk properties of the  system show abrupt changes for any $\epsilon \neq 0$. 
In particular, a significant result in the present paper is that the fraction of surviving walker shows an inverse logarithmic decay
for $\epsilon < 0$. 
Usually we find a power law decay in one dimension  with possibly a logarithmic correction, e.g., in 
\cite{dand} and purely logarithmic in rare cases, an example in higher than two dimensions can be found in \cite{bennaim}.  

For $\epsilon>0$ the bulk properties (e.g., persistence probability, fraction of surviving particles) 
show  universality in the sense
there is a  unique 
scaling behaviour of the dynamical quantities independent of $\epsilon$.
As a negative bias is incorporated in the system, 
%not only the microscopic properties 
 both the fraction of surviving particles and persistence probability   show a $\epsilon$ dependent behaviour.   
The persistence probability also does not  show a power law dependence on time. The behaviour of the bulk properties 
 can be qualitatively 
understood; the nature of the bias makes the particles more long lived and as a consequence, the probability of a site remaining
unvisited decays faster than a power law.   
% for . For $\epsilon<0$,
% microscopic properties depends on $\epsilon$ in a constant manner.

At the microscopic level the system also shows completely different behaviour for 
positive and negative bias. First, the distributions have a different nature (Gaussian, single peaked) and also show a $\epsilon$ dependent scaling behaviour for the 
negative bias. Secondly, there is no crossover behaviour as found for the  positive bias case.
The negative bias case is entirely dominated by the repulsion from an early stage which causes rapid change of 
direction such that  $S(t)$  increases as $\epsilon$ becomes more negative. 
%This also explains why the 

%persistence probability has a fast decay; as annihilations are less probable, the probability of the sites being visited by one of the particles

%increases.  

 For the fully biased case, $\epsilon=\epsilon_c=-0.5$, the motion is effectively the same as the repulsive motion
 between random walkers \cite{arratia} where the scaling behaviour is known to be $x \sim t^{1/4}$,
which is also obtained from the simulations. 
Here we find in general  $x \sim t^{\alpha}$;
an interesting issue is the dependence of $\alpha$ on $\epsilon$. 
The present results suggest  that  $\alpha$ has a weak dependence on $\epsilon$ for $\epsilon < -0.1$; it continuously
decreases from $\sim 0.3$ to $0.25$ for $-0.1 \geq \epsilon \geq -0.5$.
This has been confirmed using two different methods. 
We also find that  $\alpha - \alpha_c$ ($\alpha_c$ corresponds to $\epsilon_c$) increases in a power law manner with $\epsilon - \epsilon_c$. The fact that we get $\alpha_c \simeq 0.244$ and not exactly 0.25 from Method II possibly indicates the presence of   a finite  size effect.  
On the other hand there is a sharp decay in the value of $\alpha$ from 0.5 to $\sim 0.3$ as $\epsilon$ deviates from zero. 

%The results being numerical, the possibilities  that (a)  $\alpha$ becomes 1/4 for any $\epsilon < 0$ or (b) 
%$\alpha$ has two distinct values $\sim 0.3$ and 0.25 for $\epsilon < -0.1$ are  not  excluded and we would like to keep
%this question open and subject to further investigations.
%The inter-particle distance can be 

As already mentioned, for  $\epsilon > 0$, the exponents  are independent of $\epsilon$ while for $\epsilon < 0$, there is a non-universality. 
The former case  is comparable to a 
 system of charge-less massive  particles with a variable gravitational interaction between nearest  
 neighbours, the variation arising from the diffusive component. For the latter case when $\epsilon< 0$, the system    resembles  a collection of 
like charges with variable Coulomb interaction, the diffusive component 
again responsible for the variation.  
Of course, the annihilation factor is present in both cases  such that a 
simple mapping to a  system with gravitational or Coulomb interaction is not 
sufficient. 
For  $\epsilon < 0$, in the extreme limit of $\epsilon = \epsilon_c$, 
the diffusive component is absent and these charged particles may be regarded as electrons in a lattice  perturbed from their equilibrium positions resulting in the well known oscillatory behaviour. This is because the particles 
attain a equidistant configuration at later times and the movements may be regarded as perturbations about their equilibrium positions. 

 For positive $\epsilon$, the diffusive component does not affect the exponents and only  causes a crossover 
behaviour while for $\epsilon < 0$, the diffusive component is 
more relevant as both the bulk and tagged particle 
dynamics show strong $\epsilon$ dependence. Hence, in a way, the gravitational interaction appears to be more `robust' in comparison. The reason may be related to the fact that the annihilation factor is more effective for $\epsilon >0$; for $\epsilon < 0$, cases where the neighbours are equidistant occur more frequently, thereby   enhancing the diffusive factor. This is evident from the snapshots even  when $\epsilon$ is small in magnitude.

%domain growth phenomenon in zero

%temperature quench of Ising model under Glauber spin flip dynamics.

%At large time when each particle become almost equidistant from their neighbouring particles 
%on the lattice, they perform a nearly  oscillatory 
%motion for $\epsilon=-0.5$. Hence, the particles are not static and an active steady state 
%is achieved. The present study is able to explain the
%role of the bias in the macroscopic level as well as the individual level. 

Acknowledgement: The authors thank DST-SERB project, File no. EMR/2016/005429 
(Government of India) for financial support. Discussion with Purusattam Ray is also acknowledged.

\section{ORCID iDs}
Reshmi Roy https://orcid.org/0000-0001-6922-6858 \\
Parongama Sen https://orcid.org/0000-0002-4641-022X

\end{document}